\documentclass{iopart}
\usepackage{epsfig}

\begin{document}

\jl{4}

\vspace*{-2.5cm}
\title{Hadronic centrality dependence in nuclear collisions}
\author{ Sonja Kabana }
\address{Laboratory for High Energy Physics, University of Bern,
    Sidlerstrasse 5, 3012 Bern, Switzerland, e-mail: sonja.kabana@cern.ch}
\begin{center}
\begin{abstract}
\noindent
The kaon number density in nucleus+nucleus and p+p reactions is investigated
for the first time
as a function of the initial energy density $\epsilon$
 and is found to exhibit
 a  discontinuity around $\epsilon$=1.3 GeV/fm$^3$.
This suggests a higher degree of chemical equilibrium for $\epsilon >$
 1.3 GeV/fm$^3$.
It can also be interpreted as reflection of the same
discontinuity, appearing in the chemical 
freeze out temperature (T) as a function of
$\epsilon$.
The  $N^{\alpha \sim 1}$ dependence of (u,d,s) hadrons, whith N the
number of participating nucleons,
also indicates a high degree of chemical equilibrium and T saturation,
 reached
at $\epsilon >$1.3 GeV/fm$^3$.
Assuming that 
the 
 intermediate mass region (IMR) 
dimuon enhancement seen by NA50 is due to open
charm ($D \overline{D}$), the following observation can be made:
a) Charm is not equilibrated.
b) $J/\Psi/D \overline{D}$ suppression -unlike $J/\Psi/DY$-
appears also in S+A collisions, above $\epsilon$ $\sim$1 GeV/fm$^3$.
c) Both charm and strangeness show a discontinuity near the same $\epsilon$.
d) $J/\Psi$ could be formed mainly through $c \overline{c}$ coalescence.
e) The enhancement factors of hadrons with u,d,s,c quarks
may be connected in a simple way to the mass gain of these particles
if they are produced out of a quark gluon plasma (QGP).
We discuss these results as possible evidence for the QCD phase transition
occuring near $\epsilon \sim $1.3 GeV/fm$^3$.

\end{abstract}
\end{center}

\section{Introduction}
\noindent
The quark-gluon plasma phase transition predicted by QCD \cite{qcd}
 may occur and manifest itself in ultrarelativistic nuclear collisions
 through discontinuities in the 
initial energy density ($\epsilon_i$) dependence of relevant observables.
A major example of a discontinuity is seen in the
 $J/\Psi/DY$ \cite{na50} discussed e.g. in \cite{satz_review,pbm}.
We investigate here for the first time the dependence of 
strangeness production, in particular of kaons, on the initial energy density
$\epsilon_i$.
The degree of equilibrium achieved in nuclear collisions has been
intensively studied comparing hadron ratios and densities to models
(see e.g. \cite{pbm,redlich,becatini,rafelski}).
We investigate here if chemical equilibrium is achieved,
examining an other aspect of equilibrium states,
namely the volume ($V$) independence of hadron densities
($\rho$).
\begin{figure}[t]
\begin{center}
\epsfig{figure=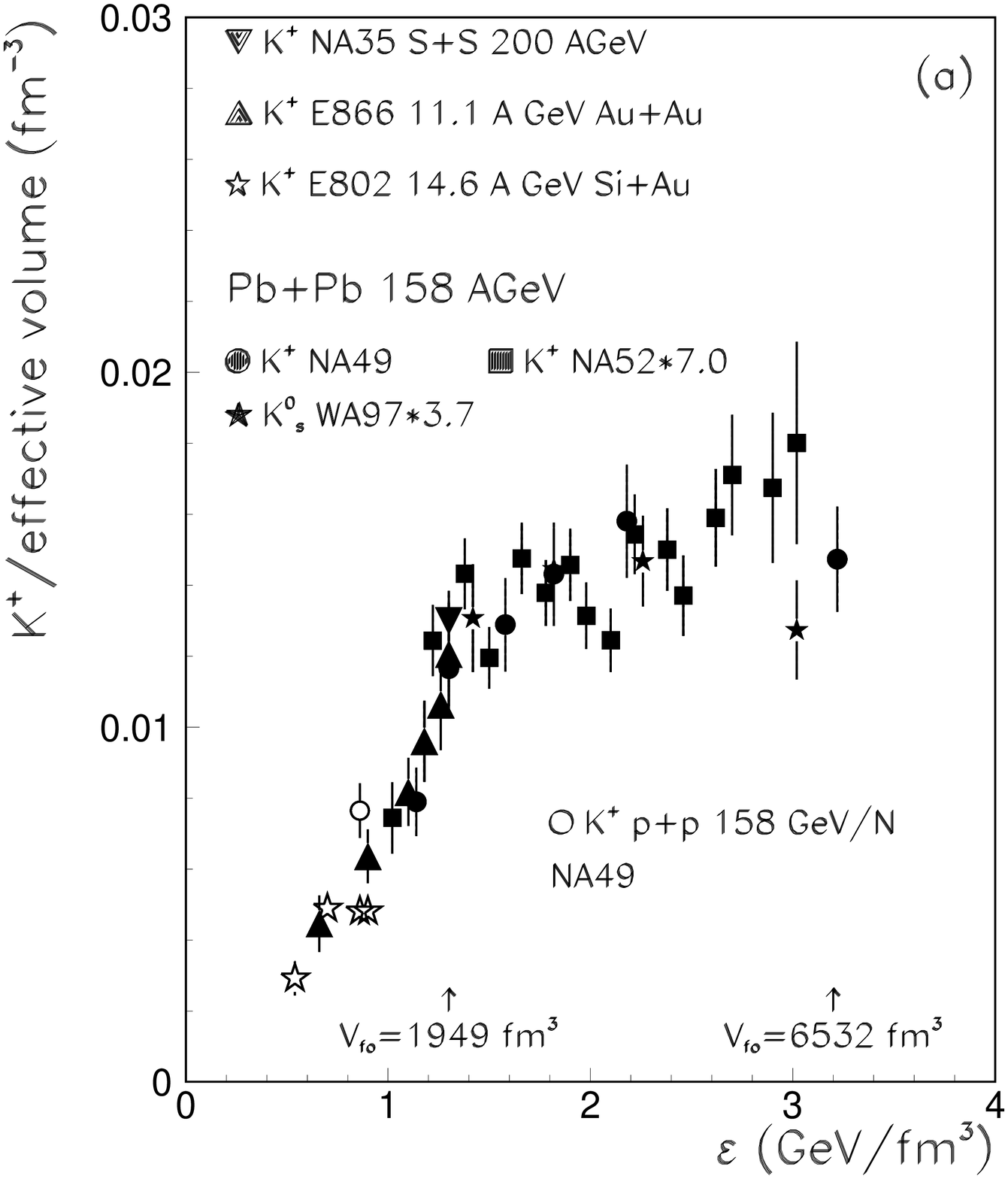, width=60mm, height=55mm}
\epsfig{figure=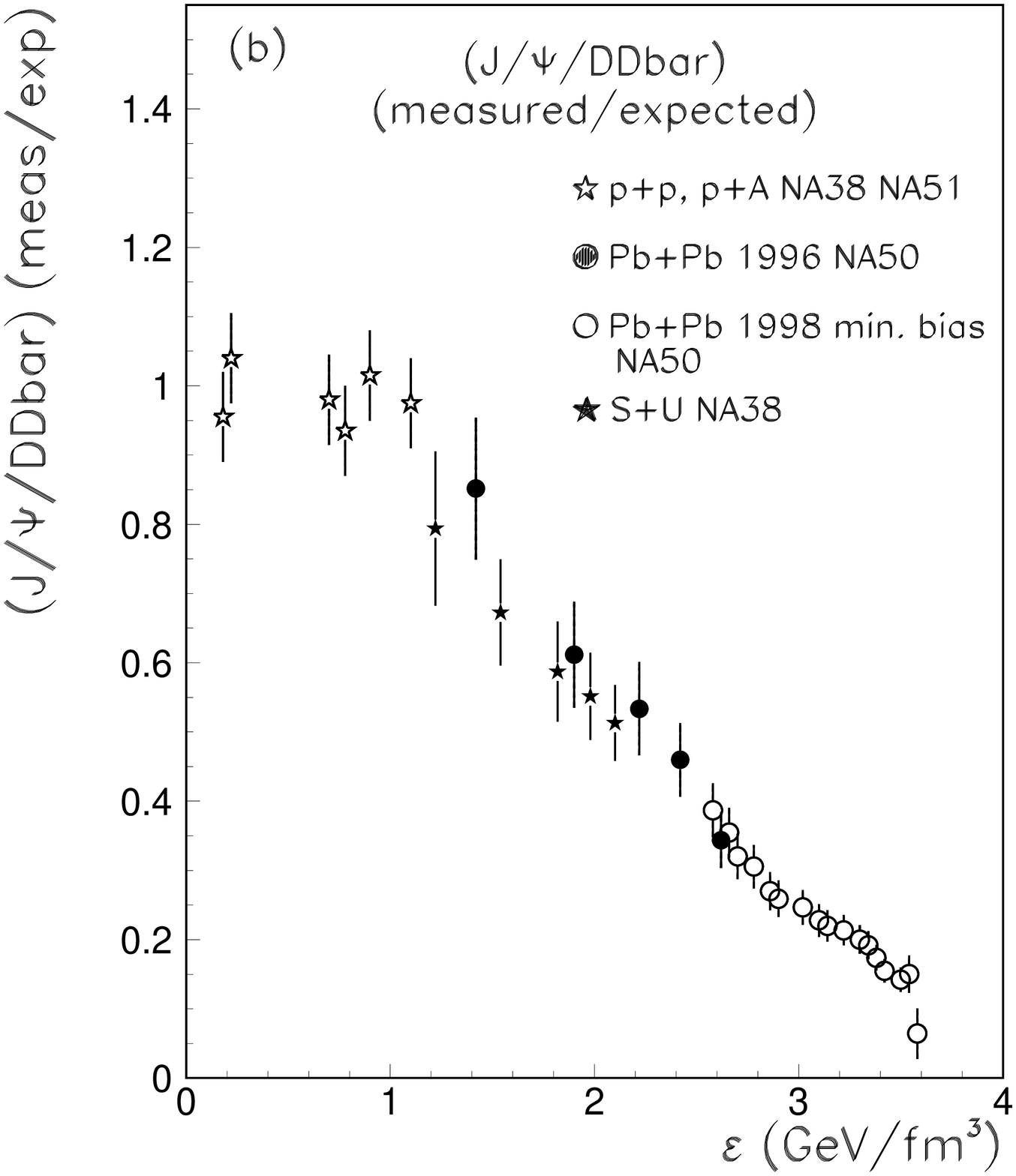, width=60mm, height=55mm}
\caption{Initial energy density ($\epsilon$) dependence of:
(a) The  $K^+$ multiplicity over the effective volume 
of the particle source at thermal freeze out.
(b) The $J/\Psi/D \overline{D}$ (measured/'expected') ratio
\protect\cite{0004138}.
}
\end{center}
\label{kaons}
\end{figure}
\begin{figure}[t]
\begin{center}
 \epsfig{figure=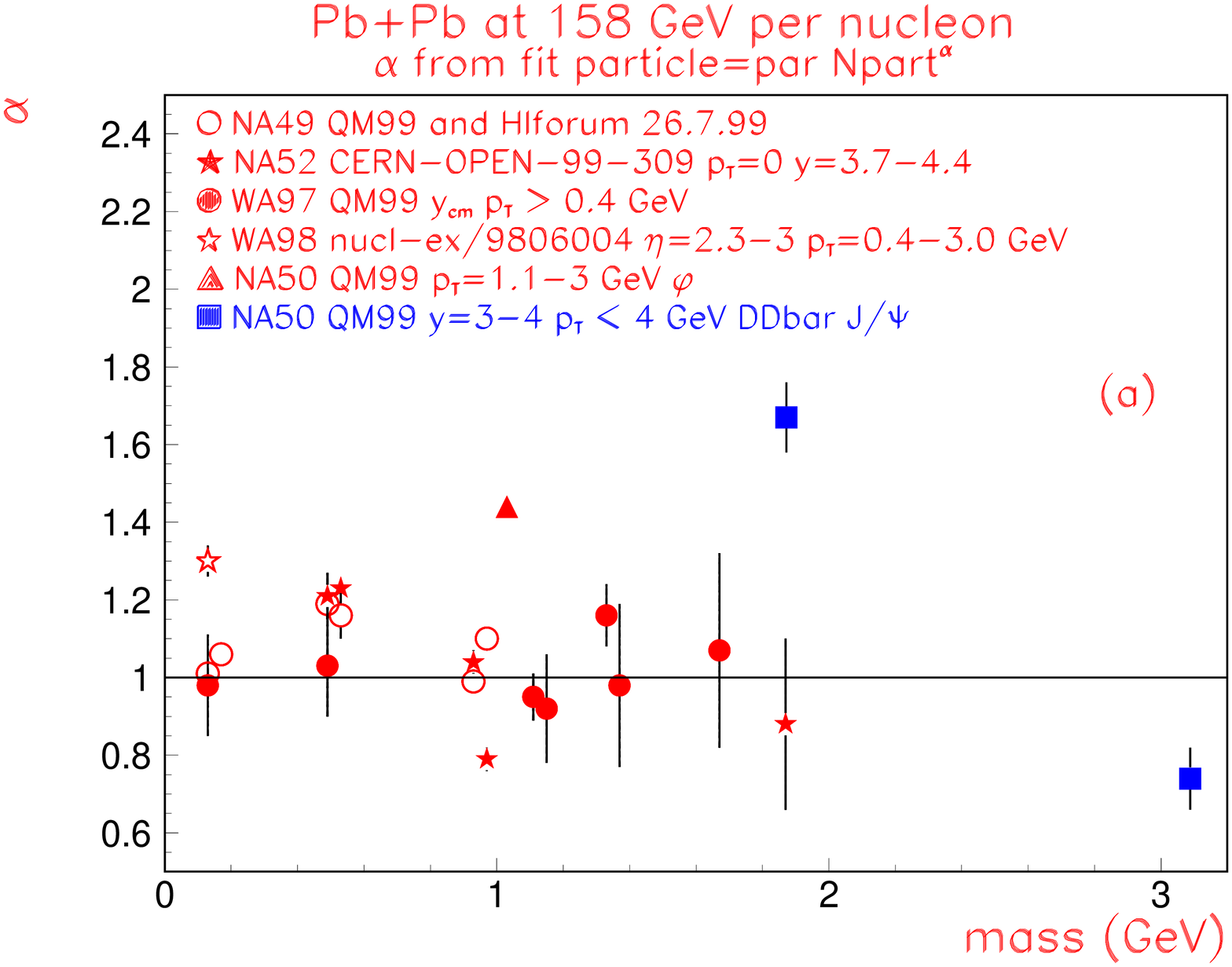, width=60mm, height=55mm}
\epsfig{figure=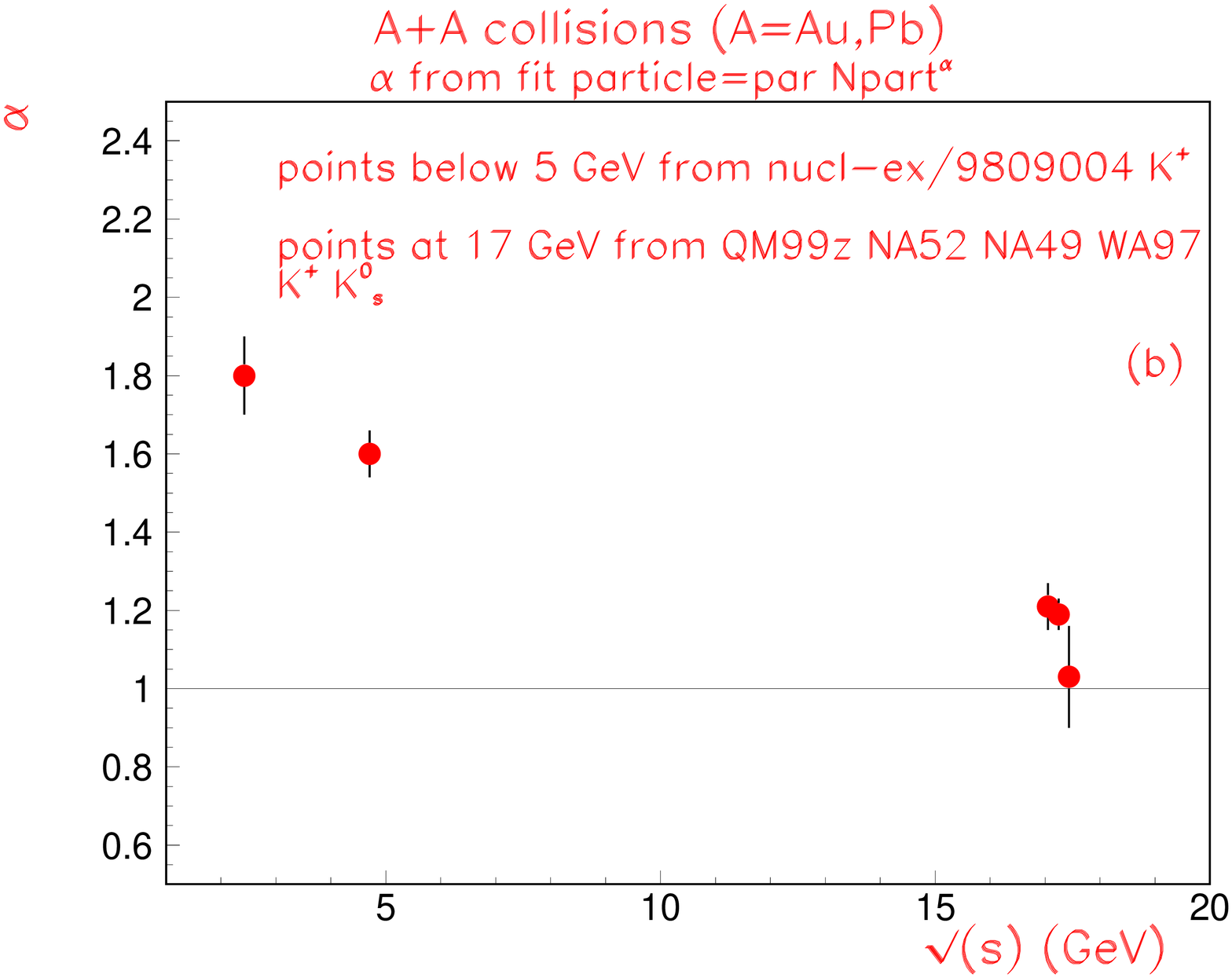, width=60mm, height=55mm}
\caption{
The parameter $\alpha$, resulting from the
$N^{\alpha}$ fit to hadron yields shown as a function of 
(a): the mass of the particles in the region $\epsilon>$1.3GeV/fm$^3$ at SPS
and
(b) of the $\sqrt{s}$, for kaons.
N is the number of participating nucleons.
}
\end{center}
\label{kaons}
\end{figure}
\section{Results and discussion}
\noindent
The kaon density ($\rho_K$=(K per collision)/V) at the thermal freeze out
in nuclear reactions, investigated 
as a function of the initial energy density $\epsilon_i$ (figure 1,(a)) 
(see \cite{0004138} for calculation details),
exhibits a dramatic changeover around $\epsilon$=1.3 GeV/fm$^3$,  saturating
 for higher $\epsilon$ values, while it is falling below.
The syst. error on $\epsilon_i$ is estimated to be $\sim$ 30\%.
It is assumed that the number of
nucleons participating in the collision (N) is proportional
to the volume of the particle source at the thermal freeze out
\cite{0004138}.
The new results from Si+Au at 14.6 A GeV and p+p at 158 A GeV 
shown in figure 1, which are not included in \cite{0004138}, 
have been estimated using data  from
\cite{siau} and methods described in \cite{0004138}.
Furthermore, $\rho_K$ rises with N respectively with V below
$\epsilon$=1.3 GeV/fm$^3$ 
while it does not depend on N respectively on V above $\epsilon$=1.3 GeV/fm$^3$.
To illustrate this, two values of $V$ are noted on figure 1.
The changes of $K^{\pm}$ and $\pi^{\pm}$ with N within the Pb+Pb system, 
have been first realized in \cite{na52_centr}.
A similar behaviour as the one seen in figure 1, can be inferred for
 pions as well as for the $K/\pi$ ratio (S.K. work in progress).
\noindent
The $N^{\alpha}$ exponent of hadrons with (u,d,s) quarks
above $\epsilon$=1.3 GeV/fm$^3$,
do not depend on the particle mass (figure 2, (a)).
At $\epsilon>$1.3 GeV/fm$^3$ $\alpha$ is near to one,
as expected in case of a chemically equilibrated
state, assuming N $\sim$ V.
 The deviations seen in $\phi$, $\pi^0$ and $\overline{p}$ may be 
 due to the transverse momentum acceptance.
Therefore, figure 2 (a) supports the assumption of
 a high degree of chemical equilibrium  reached 
above $\epsilon$=1.3 GeV/fm$^3$, among hadrons with u,d,s quarks.
The
 $N^{\alpha}$ exponent of kaons
 is found to depend strongly on $\sqrt{s}$ for kaons (figure 2,(b)).
Therefore, below $\epsilon$=1.3 GeV/fm$^3$, $\rho_k$ 
(figure 1 and figure 2 (b)), 
$\rho_{\pi}$ and the $K/\pi$ ratio,
 show an increase with increasing $N$ respectively with $V$.

\noindent
Figures 1 and 2 can be interpreted in two ways.
Firstly,
 kaons may achieve a higher degree of chemical equilibrium 
only 
for $\epsilon >$ 1.3 GeV/fm$^3$, and may not be fully equilibrated
 below \cite{0004138}.
The equilibration of strangeness is expected in a QGP 
and its observation at $\epsilon \sim $ 1.3 GeV/fm$^3$
 could therefore be a sign of a transition to QGP.
In this case, 
 it is a transition from a non equilibrated
hadron gas to an equilibrated QGP.
 It is therefore not a well defined phase
transition in the thermodynamic sense.
\\
\noindent
Secondly,
 kaons can be in fact chemically equilibrated
also below  $\epsilon$= 1.3 GeV/fm$^3$, and the 
change respectively the constancy  of
$\rho_K$ with $V_{fo}$ and $\epsilon_i$ observed in figure 1, 
can be a result of the increase of the freeze out temperature with $\epsilon_i$
below $\epsilon$= 1.3 GeV/fm$^3$, respectively of
 the stability of $T_{fo}$ above 1.3 GeV/fm$^3$.
This dependence of $T_{fo}$ on $\epsilon_i$, namely rising until
it reaches a critical $T_c$ value and saturating above for all reactions,
would strongly support the QCD phase transition appearing 
at $\epsilon \sim$ 1.3 GeV/fm$^3$. 
\noindent
This interpretation fully agrees
 with thermal models  which suggest that particle ratios at freeze out
are compatible with thermalization
even in A+A collisions at 1 A GeV \cite{redlich}.
However the first interpretation is not in gross disagreement with
\cite{redlich}, because there  the thermal model description is
modified (introducing e.g. $\rho_k \sim V$)
 in order to describe the data at 1 A GeV.

\noindent
Furthermore, the correct interpretation can be corroborated by
further investigations discussed in the following.
The nonzero baryochemical potential ($\mu_B$), which in the reactions shown
in figure 1, happens to
change with $\epsilon_i$, makes the intepretation of figure 1 difficult.
Therefore, it appears that the
dependence of the temperature at chemical freeze out extrapolated to $\mu_b$=0,
 on $\epsilon_i$, would help to identify and prove the QCD phase transition. 
A rising and then a for ever saturating freeze out temperature
above $\epsilon$ 1.3 GeV/fm$^3$ is a strong argument
that the QCD phase transition occurs at this $\epsilon$,
and figure 1 is a direct consequence of it.

\noindent
The question if the QCD phase transition appears at the critical $\epsilon_i$
in any volume,
or if there is additionally a 
 critical initial
 volume of the particle source above which the transition takes place, can be answered
 comparing QGP signatures in systems with different volumes but the
same $\epsilon_i$.
For example comparing $p+p$, $e^+ e^-$ etc  collisions
to heavy ion collisions e.g. at the same $\epsilon$. 
This is not yet done for the signature of the $J/\Psi$ suppression 
and it has to be clarified e.g. using Tevatron data \cite{0004138}.
For the signature of strangeness enhancement 
it is suggested by figure 1 in \cite{becatini}
that there is indeed a critical initial volume,
 only above which strangeness is enhanced over $p+ \overline{p}$ 
{\it at the same $\epsilon_i$}.
This conclusion follows, if we assume that Tevatron reaches at least
$\epsilon_i$ values similar to SPS A+A collisions \cite{na35}
and if figure 1 in \cite{becatini} is not 
biased by the model calculation \cite{becatini}.

\noindent
If strangeness is indeed not equilibrated at $\epsilon < 1.3$ GeV/fm$^3$,
this may explain 
 the decrease of the double ratio ($K/\pi$)(A+A/p+p) with increasing
$\sqrt{s}$.
In particular, a larger strangeness annihilation is enforced by equilibrium
at SPS reducing the strange particle yield.
However the assumption of 
non equilibrium of $s \overline{s}$ at low $\epsilon$
 is not nessecary here, since the above observation
can be possibly traced back
 to e.g. the variation of $\mu_B(A+A) / \mu_B(p+p)$ with $\sqrt{s}$
in A+A collisions.
Furthermore,
in the context of QGP formation, it seems irrelevant to discuss
e.g.  $s \overline{s}$ enhancement in A+B over p+p collisions
 in a nonequilibrium situation.
It is the very establishment of equilibrium in the (u,d,s) sector, 
 which can reveal informations on QGP.
\\
The kaon number densities in p+p and A+B collisions
in figure 1, (a) are similar, when compared at the same $\epsilon_i$.
See also \cite{heinz} for a discussion of universality of
 pion phase space densities.

\noindent
Our prediction for the N dependence of
hadrons at RHIC and LHC is the $N^1$ thermal limit,
as long as hadron yields are 
 dominated by low transverse momentum  particles.
Furthermore,
if the changeover of $\rho_k$ at $\epsilon=$1.3 GeV/fm$^3$ shown in
figure 1 is due to the QCD phase transition,
we predict for RHIC and LHC the same 
total strangeness (or kaon) number density and the
same freeze out temperature, -after correction for the $\mu_B$ dependence-,
as for $\epsilon=$1.3-3.0 GeV/fm$^3$.
If this change is
 however due to the onset of equilibrium in a hadronic gas, 
and the QCD phase transition takes place at 
higher $\epsilon$, it may manifest itself through a 
second changeover of hadron number densities, 
ratios and freeze out temperatures -after correction for the different
$\mu_B$-  e.g. in RHIC
above  $\epsilon $ $\sim$ 3 GeV/fm$^3$.

\noindent
Assuming that the IMR dimuon enhancement seen by NA50 is due to open
charm, the following observations can be made:
a)
open charm appears not to be equilibrated ($\alpha=1.7$) (figure 2, (a))
 \cite{0004138}.
b) The $J/\Psi/D \overline{D}$ ratio deviates from p+p and p+A data
also in S+U collisions (figure 1, (b)), above $\epsilon$ $\sim$1 GeV/fm$^3$.
c) It therefore appears that both charm and strangeness show a discontinuity
near the same $\epsilon$$\sim$1 GeV/fm$^3$  \cite{0004138},
 similar to the critical $\epsilon_c \sim$1-2 GeV/fm$^3$ predicted by QCD
\cite{qcd,satz_review}.
d) The N dependence of the $J/\Psi/D \overline{D}$ ratio 
can be interpreted as the $J/\Psi$ being 
formed through $c, \overline{c}$ coalescence \cite{0004138}.
e) Finally, the enhancement factors of hadrons with u,d,s,c quarks
may be connected in a simple way to the mass gain of these particles
in the quark gluon plasma (table below) \cite{skpm}.
$T_q$ are the enhancement factors of the lightest mesons with
u,d,s,c quarks ($\pi, K, D$), if they are produced out of a quark gluon 
plasma (e.g. $g + g \rightarrow s + \overline{s} $ (1)),
 as compared to their direct production from hadron interactions
away from the transition point
(e.g. $p + p \rightarrow K^+ + \Lambda +p$ (2)). 
The gain is taken proportional to $m_{particle} - m_{quarks}$, as this
expresses the different thresholds of reactions (1) and (2).
In the table below
 the predicted enhancement factors ($T_q$) of hadrons with u,d,s,c quarks 
from a QGP are compared 
to the experimentally measured ones ($E_q$), and are found to be similar.
(Definitions:
$th_q=m_0 - m_q$, $m_{u,d}$=7 MeV, $m_s$=175 MeV, $m_c$=1.25 GeV,
 $m_0=m(\pi, K, D)$).
\begin{table}[ht]
\begin{tabular}{lllll}
Quark flavour  & $th_q$ & $T_q=\sqrt{ th_q / th_{u,d}}$ & $E \frac {(A+B} {(N+N)} $ & $E_q=E/E_{u,d} $
\\
\hline
 u,d  & 133 & 1 &   $\pi/N$ $\sim$ 1.12 & 1
\\
 s    & 320 & 1.55 &  $K/N$ $\sim$ 2   & 1.79
\\
 c    & 615 & 2.15  & $D \overline{D}$ meas/exp $\sim$ 3 & 2.68
\\
\end{tabular}
\end{table}
\\
\noindent
{\bf Acknowledgments}
I would like to thank Prof.~K~Pretzl and Prof.~P~Minkowski for 
stimulating discussions and 
  the Schweizerischer Nationalfonds for their support.

\end{document}